\documentclass{eptcs}
\usepackage{amsmath}
\usepackage{amsmath,amssymb,oldlfont}
\usepackage{proof}
\usepackage{breakurl}

\newcommand{\out}[1]{}
\newcommand{\delete}[1]{}
\newcommand{\blank}[1]{}
\newcommand{\notedomission}[1]{\medskip\noindent{\bf TEXT OMITTED}\\[2mm]}

\newcommand{\bxit}[1]{\hbox{\it #1}}

\newcommand{\bxrm}[1]{\hbox{\rm #1}}

\newcommand{\bxbf}[1]{\hbox{\bf #1}}

\newcommand{\bxsc}[1]{\hbox{\sc #1}}
\newcommand{\bxtt}[1]{\hbox{$\tt #1$}}

\newcommand{\fnc}[1]{\mbox{\underline{\it #1}$\,$}}

\newcommand{\bfa}{{\bf a}}
\newcommand{\bfb}{{\bf b}}
\newcommand{\bfc}{{\bf c}}
\newcommand{\bfd}{{\bf d}}

\newcommand{\bff}{{\bf f}}

\newcommand{\bfq}{{\bf q}}

\newcommand{\bfs}{{\bf s}}
\newcommand{\bft}{{\bf t}}

\newcommand{\itp}{{\it p}}

\newcommand{\tta}{\hbox{$\tt a$}}
\newcommand{\ttb}{\hbox{$\tt b$}}
\newcommand{\ttc}{\hbox{$\tt c$}}

\newcommand{\ttf}{\hbox{$\tt f$}}
\newcommand{\ttg}{\hbox{$\tt g$}}

\newcommand{\tts}{\hbox{$\tt s$}}
\newcommand{\ttt}{\hbox{$\tt t$}}

\newcommand{\ttv}{\hbox{$\tt v$}}

\newcommand{\ttzero}{\hbox{$\tt 0$}}

\newcommand{\ttone}{\hbox{$\tt 1$}}

\newcommand{\bfP}{{\bf P}}

\newcommand{\bfzero}{{\bf 0}}
\newcommand{\bfone}{{\bf 1}}

\newcommand{\calA}{\hbox{$\cal A$}}

\newcommand{\calC}{\hbox{$\cal C$}}
\newcommand{\calD}{\hbox{$\cal D$}}
\newcommand{\calE}{\hbox{$\cal E$}}
\newcommand{\calF}{\hbox{$\cal F$}}

\newcommand{\calP}{\hbox{$\cal P$}}

\newcommand{\calS}{\hbox{$\cal S$}}

\newcommand{\calX}{\hbox{$\cal X$}}

\newcommand{\ttN}{\hbox{$\tt N$}}

\newcommand{\dN}{\hbox{$\Bbb N$}}

\newcommand{\gra}{\hbox{$\alpha$}}

\newcommand{\grd}{\hbox{$\delta$}}

\newcommand{\grh}{\hbox{$\eta$}}

\newcommand{\grl}{\hbox{$\lambda$}}
\newcommand{\grm}{\hbox{$\mu$}}
\newcommand{\grn}{\hbox{$\nu$}}

\newcommand{\grp}{\hbox{$\pi$}}

\newcommand{\grs}{\hbox{$\sigma$}}
\newcommand{\grt}{\hbox{$\tau$}}

\newcommand{\grf}{\hbox{$\varphi$}}

\newcommand{\grq}{\hbox{$\psi$}}
\newcommand{\grw}{\hbox{$\omega$}}

\newcommand{\grG}{\hbox{$\Gamma$}}

\newcommand{\grX}{\hbox{$\Xi$}}
\newcommand{\grP}{\hbox{$\Pi$}}

\newcommand{\hd}{\hbox{ {\tt hd} }}
\newcommand{\tl}{\hbox{ {\tt tl} }}

\newcommand{\ra}{\rightarrow}
\newcommand{\pa}{\rightharpoonup}
\newcommand{\sra}{\!\rightarrow\!} 

\newcommand{\splus}{\!+\!}

\newcommand{\arity}{\fnc{arity}}

\newcommand{\qed}{\hfill {\Large\boldmath$\Box$}}
\newcommand{\lng}{\langle}
\newcommand{\rng}{\rangle}
\newcommand{\df}{=_{\rm df}}

\newcommand{\rsem}{]\hspace{-0.5mm}]} 
\newcommand{\lsem}{[\hspace{-0.5mm}[} 

\newcommand{\ignore}[1]{}

\newcommand{\zero}{\rule{0mm}{3mm}}

\newcommand{\bc}{\begin{center}}
\newcommand{\ec}{\end{center}}
\newcommand{\beq}{\begin{equation}}
\newcommand{\eeq}{\end{equation}}
\newcommand{\be}{\begin{enumerate}}
\newcommand{\ee}{\end{enumerate}}
\newcommand{\bi}{\begin{itemize}}
\newcommand{\ei}{\end{itemize}}
\newcommand{\bd}{\begin{description}}
\newcommand{\ed}{\end{description}}
\newcommand{\beqn}{\begin{equation}}
\newcommand{\eeqn}{\end{equation}}

\newcommand{\beqna}{\begin{eqnarray}}
\newcommand{\eeqna}{\end{eqnarray}}
\newcommand{\beqnas}{\begin{eqnarray*}}
\newcommand{\eeqnas}{\end{eqnarray*}}
\newcommand{\beqnaa}{$$\begin{array}{rcll}}  
\newcommand{\eeqnaa}{\end{array}$$}  
\newcommand{\beqnana}{$$\begin{array}{lrcll}}  
\newcommand{\eeqnana}{\end{array}$$}  
\newcommand{\btbl}[1]{\begin{center}\begin{tabular}{#1}}
\newcommand{\etbl}{\end{tabular}\end{center}}
\newcommand{\beqnc}{$$\begin{array}{rclcl}}
\newcommand{\eeqnc}{\end{array}$$}
\newcommand{\fn}{\footnote}

\newcommand{\prf}{{\bf Proof. }}

\newtheorem{dclprop}{{\sc Proposition}} 
\newtheorem{dclprops}{{\sc Proposition}}[subsection] 
\newtheorem{dclbigthm}[dclprop]{THEOREM}

\makeatletter
\def\thmlabel#1{\@bsphack\if@filesw {\let\thepage\relax
\xdef\@gtempa{\write\@auxout{\string
\newlabel{#1}{{\@Roman{\@currentlabel}}{\thepage}}}}}\@gtempa
\if@nobreak \ifvmode\nobreak\fi\fi\fi\@esphack}
\makeatother

\newtheorem{dclthm}[dclprop]{{\sc Theorem}}   
\newtheorem{dclthms}[dclprops]{{\sc Theorem}}   

\newtheorem{dcllem}[dclprop]{{\sc Lemma}}
\newtheorem{dcllems}[dclprops]{{\sc Lemma}} 
\newtheorem{dclsublem}[dclprop]{{\sc Sublemma}}
\newtheorem{dclcor}[dclprop]{{\sc Corollary}}
\newtheorem{dclcors}[dclprops]{{\sc Corollary}} 
\newtheorem{dcldfn}[dclprop]{{\sc Definition}}
\newtheorem{dcldfns}[dclprops]{{\sc Definition}}
\newtheorem{dclasss}[dclprops]{{\bf Assumption}}
\newtheorem{dclass}[dclprop]{{\bf Assumption}}

\newenvironment{prop}{\medskip\begin{dclprop}\sl}{\end{dclprop}}

\newenvironment{thm}{\begin{dclthm}\sl}{\end{dclthm}}

\newenvironment{lem}{\medskip\begin{dcllem}\sl}{\end{dcllem}}

\newcommand{\bsl}{\begin{verse}\sl}
\newcommand{\esl}{\end{verse}}

\newtheorem{exs}[dclprop]{Examples}
\newtheorem{ex}[dclprop]{Example}

\newtheorem{exxs}[dclprop]{Exercises}

\newenvironment{exercises-with-preamble}{\begin{exxs}\rm}{\end{exxs}}

\newcommand{\bz}{\begin{quote}\small}
\newcommand{\ez}{\end{quote}}

\newcommand{\einference}[2]  
  {\shortstack
      {$ #1 $\\ \mbox{}\\ $ #2 $}}

\newcommand{\nsubsection}[1]{\setcounter{subsubsection}{0}\addtocounter{subsection}{1} \smallskip\bigskip\noindent{\large\bf \arabic{section}.\arabic{subsection} \ #1}\medskip}

\newlength{\txtlth}
\newlength{\txtht}

\newcommand{\cm}{\bxit{Cmp}}
\newcommand{\dc}{\bxit{Dcm}}

\title{Implicit complexity for coinductive data:\\
	a characterization of corecurrence}
\author{Daniel Leivant \institute{Indiana University and Loria Nancy} 
	\email{leivant@indiana.edu}
	\and
	Ramyaa Ramyaa \institute{Indiana University and Universitat Munchen}
	\email{ramyaa@indiana.edu}
}
\def\titlerunning{Implicit corecurrence}
\def\authorrunning{Leivant and Ramyaa}

\begin{document}
\maketitle

\begin{abstract}
We propose a framework for reasoning about programs that
manipulate coinductive data as well as inductive data.
Our approach is based on using equational programs, which support a seamless
combination of computation and reasoning, and using 
productivity (fairness) as the fundamental assertion, rather than 
bi-simulation.  The latter is expressible in terms of the
former.

As an application to this framework, we give an implicit characterization 
of corecurrence: a function is definable using 
corecurrence iff its productivity is provable
using coinduction for formulas in which 
data-predicates do not occur negatively.
This is an analog, albeit in weaker form, of 
a characterization of recurrence (i.e.\ primitive
recursion) in \cite{Leivant-unipolar}. 
\end{abstract}

\section{Introduction}

Coinductive data has been recognized for nearly two decades as a 
powerful framework for dealing with infinite objects of evolving and
computational nature, such as streams, and --- more generally --- 
the behavior of unbounded processes and dynamic systems.  

We consider computation over
``data-systems", in which data-types may be defined both
inductively and co-inductively.
As our main computation model we use equational programs,
since these have immediate kinship with
formal theories: a program's equations can be viewed as axioms,
and computations are simply derivations in equational logic.
In the first part of this paper we develop some building blocks for this
project.  We consider the {\em global} semantics of
programs $P$ over a data-system, that is their behavior as
``uninterpreted programs" over all
structures for the vocabulary of the data-system. 
This approach was developed for {\em inductive} data in \cite{Leivant-intrinsic};
here we extend it to data-systems in general, including coinductive
constructions.
It is orthogonal to category theoretical methods
in the study of coinduction, which seek to characterize the intended (canonical)
model. 


An important benefit of streamlined proof systems for reasoning about programs
is their use for characterizing major computational complexity
classes.  Such characterizations fall within the realm of {\em implicit 
computational complexity,} where one delineates complexity
classes without reference to computational resources such as time and
space.  In particular, there are illuminating characterizations
of complexity classes in terms of the strength of proof 
methods needed to prove
termination (see e.g.\ \cite{Buss86,Leivant-foundational,Leivant-unipolar}).
Such results lend insight into
the significance of complexity classes, provide natural frameworks 
for programming within given complexity boundaries, 
and yield static analysis tools for guaranteeing complexity.
Implicit characterizations have further potential benefit 
for coinductive data, because they
might clarify complexity notions
that are dual to traditional notions of computational complexity such as
Polynomial Time.

The primitive recursive functions over the set \dN\ of natural 
numbers were characterized proof theoretically already by Parsons \cite{Parsons70},
who proved that a function is primitive recursive iff it is
provable in Peano's Arithmetic with
induction restricted to existential formulas.

In \cite{Leivant-LCC,Leivant-intrinsic} we developed {\em intrinsic theories,}
a generic framework for reasoning about equational computing over inductive data, and in
\cite{Leivant-unipolar} we used it to characterize the primitive recursive
functions in terms of induction for a particular class of formulas.
Call a formula {\em unipolar} if it does not use data-predicates
(i.e.\ references to data) in both
positive and negative position; an example are the {\em positive} formulas,
in which data-predicates do not occur in a negative position.
In \cite{Leivant-unipolar} we proved that a computable function is primitive recursive
iff it is provably correct in the intrinsic theory for \dN\ with induction restricted
to unipolar formulas.  In fact we proved more.
The forward implication can refer to a very weak formalism, namely,
every primitive recursive function is provable, using minimal logic,
by induction for formulas in which data-predicates appear only strictly-positively.\fn{Recall
that \grf\ is a strictly-positive subformula of \grq\ if \grf\ is not in the
scope of a negation or the negative scope of an implication.}
On the other hand, for the backwards implication we proved that if a computable function is
provable, using classical logic, by induction on unipolar formulas, then it is primitive
recursive.

We establish here a dual characterization for coinductive data, but where both
implication refer to a weak deductive calculus:
a computable function over boolean streams is
primitive corecursive (i.e.\ definable using explicit definitions and
corecurrence) iff it is provable using minimal logic, by coinduction for formulas
built from only conjunction, disjunction, and existential quantification.
At present we do not know whether this result can be strengthen to show that
every equational program over streams which is provable, using {\em classical}
logic and {\em unipolar} coinduction is primitive-corecursive.

\section{Equational programs over data systems}

\subsection{Equational programs}

We describe a generic framework for
data-types that are defined using induction, coinduction, or a mix thereof.
Such frameworks are well-known for typed lambda calculi, with operators
\grm\ for smallest fixpoint and \grn\ for greatest fixpoint.  
Our present approach is to express computational
behavior of programs via global semantics, thereby dispensing with 
partial functions; and to define types semantically,
via first order axiomatics,
dispensing with explicit fixpoint operators.

A {\em constructor-vocabulary} is a finite set \calC\ of function
identifiers, referred to as {\em constructors,} each assigned an {\em arity}
$\geq 0$ (as usual, constructors of arity 0 are {\em object-identifiers}).
We posit an infinite set \calX\ of {\em variables,} and an infinite 
set \calF\ of function-identifiers, dubbed {\em program-functions,} and 
assigned arities $\geq 0$ as well. The sets \calC, \calX\ and \calF\ are,
of course, disjoint.  

If \calE\ is a set consisting
of function-identifiers and (possibly) variables, we write $\bar{\calE}$
for the set of terms containing \calE\ and closed under application:
if $g \in \calE$ is a function-identifier of arity $r$, and $t_1 \ldots t_r$
are terms, then so is $g\,t_1\, \cdots\, t_r$.  We use informally the
parenthesized notation $g(t_1, \ldots, t_r)$, when 
convenient.\footnote{In particular, when $g$ is of arity 0, it is itself a term,
whereas with parentheses we have $g()$ (with $r=0$ arguments) as a term.}
We refer to elements of $\bar{\calC}$, $\overline{\calC\cup\calX}$ and
$\overline{\calC\cup\calX\cup\calF}$ as {\em data-terms,}
{\em base-terms,} and {\em program-terms,} respectively.\footnote{Data-terms 
are often referred to as {\em values}, and base-terms as {\em patterns.}}

As in \cite{Leivant-LCC,Leivant-intrinsic}, we use an equational 
computation model, in the style of Herbrand-G\"{o}del, familiar from the
extensive literature on algebraic semantics of programs.
There are easy inter-translations between equational programs and
program-terms such as those of $\bxbf{FLR}_0$ \cite{Moschovakis89}.
We prefer to focus on equational programs because they integrate easily into 
logical calculi, and are naturally construed as mathematical theories
(with each equation as an axiom). Codifying equations by terms is, in fact,
a conceptual detour, since the computational behavior of such terms is itself 
spelled out using equations or rewrite-rules.

A {\em program-equation} is an equation of the
form $\bff(\bft_1 \ldots \bft_k) = \bfq$, where \bff\ is a program-function 
of arity $k \geq 0$, $\bft_1 \ldots \bft_k$ are base-terms, 
and $\bfq$ is a program-term.
The left-hand side of a program equation is its {\em definiendum.}
Two program-equations are {\em compatible} if their definiendums
cannot be unified.
A {\em program-body} is a finite set of pairwise-compatible 
program-equations. A program $(P,\bff)$ (of arity $k$) 
consists of a program-body $P$ and 
a program-function \bff\ (of arity $k$) dubbed the program's 
{\em principal-function.}
We identify each program with its program-body when in no danger of confusion.

We posit that every program over a given constructor-vocabulary 
has equations for destructors, as well as a discriminator.
That is, if the given vocabulary's constructors are
$\bfc_1 \ldots \bfc_k$, with $m$ the
maximal arity, then the program-functions include
the unary identifiers $\grp_{i,m}$ ($i=1..m$) and $\grd_k$,
and the program contains the equations (for
\bfc\ an $r$-ary constructor)
$$
\begin{array}{rcll}
\grp_{i,m}(\bfc(x_1, \ldots, x_r) &=& x_i & (i=1..r)\\
\grp_{i,m}(\bfc(x_1, \ldots, x_r)) 
	&=& \bfc(x_1, \ldots, x_r) & (i=r\splus 1.. m)\\[1mm]
\grd_k(\bfc_i(\vec{t}), x_1, \ldots, x_k) &=& x_i  & i = 1..k
\end{array}
$$
Thus $\grd_k$ is a definition-by-cases operation,
depending on the main constructor of the first argument.  
We call a composition of $n$ destructors $(n \geq 0)$ a {\em deep destructor}.

It is easy to define the denotational semantics of an equational program
for the canonical interpretation of inductive data.
If $(P,\bff)$ is a program for a unary function over \dN, say,
then it computes the partial function $f: \; \dN \pa \dN$
where $f(p)=q$ just in case the equation $\bff(\bar{p})=\bar{q}$
is derivable from $P$ in equational logic. (We write $\bar{n}$ for the 
$n$'th numeral, i.e.\ the data-term $\bfs\bfs \cdots \bfs \bfzero$ with $n$ 
\bfs's.

The partiality of computable functions is most commonly addressed
by either allowing partial structures 
\cite{Kleene69,AstesianoBKKMST02,Mosses04},
or by referring to domains, in which an object $\bot$ denotes divergence.
Yet another approach, adopted here, 
is based on the ``global" behavior of programs in 
all (usual, non-partial) structures.
For example, consider the program $P$ over the 
constructors $\ttzero,\tts$ consisting of the two equations\footnote{We omit
some parentheses for readability.}
$\ttf(\ttzero) = \ttzero$ and 
 $\ttf(\tts \tts x) = \ttf(\tts \tts \tts x)$.
Thus $P$ provides no instructions for input 1, and diverges for input $\geq 2$. 
The latter conditions are captured by the statement that there are structures
which model the equations $P$, and where the terms 
$\ttf(\tts \ttzero)$ and $\ttf(\tts \tts \ttzero)$ are not equal to
any numeral. 

\subsection{Global semantics}

The concept of {\em global relations,} which was present implicitly in
mathematical logic for long, came to prominence in Finite Model Theory in the
1980s. 
Let \calC\  be a collection of structures. A {\em global relation} 
(of arity $r$)
over $\cal C$ is a mapping $\cal P$ that assigns to each structure  $\cal S$
in $\cal C$ an $r$-ary relation over the universe $|\calS|$ of \calS.
For example, if \calC\ is the collection of all structures over a given
vocabulary $V$, then a first-order $V$-formula \grf, with free variables among
$x_1 \ldots x_r$, defines the predicate $\grl x_1 \ldots x_r \grf$
that to each $V$-structure \calS\
assigns the relations
$$
\{ \lng a_1 \ldots a_r \rng \mid \calS, [\vec{x} := \vec{a}] \models \grf \} 
$$
The notion that a formula delineates uniformly 
subsets of structures is implicit in 
\cite{Tarski52} and \cite{BarwiseM78}.
Alternative phrases used include {\it generalized relations}, {\it data base queries},
{\it global relations}, {\it global predicates}, {\it uniformly defined relations},
{\it predicates over oracles}, and {\it predicates}.)

A {\em global $r$-ary function} over \calC\ is defined analogously.
For example, each typed \grl-term of type $o\sra o$,  with identifiers in $V$ as primitives,
defines a global function over the class of $V$-structures.
E.g., if \ttc, \ttf\ and \ttg\ are $V$-identifiers for functions of
arity 0,1 and 2 respectively, then the term 
$\grl x,_1,x_2 \; \ttg(\ttf(x_1),\ttg(x_2,\ttc))$ defines the global function
that to each $V$-structure \calS\ assigns the mapping
$\lng x_1,x_2\rng \mapsto g(f(x_1),g(x_2,c))$, where $c,f$ and $g$ are the
interpretations in \calS\ of the identifiers $\ttc,\ttf$ and $\ttg$.  

The starting point of Descriptive
Computational Complexity \cite{Immerman89} is that programs used 
as acceptors define global relations.  When those global relations can
be defined also by certain logical formulas, one obtains machine-independent
characterizations of computational complexity classes.
For instance, 
Fagin \cite{Fagin74} and Jones \& Selman \cite{JonesS74} proved
that a predicate \calP\ over finite structures is
defined by a program running in nondeterministic polynomial time
(NP) iff it is defined
by a purely existential second order formula.

Programs of arity 0 can be used to define objects.  For example, the
singleton program $T$ consisting of the equation
$\ttt = \tts\tts\tts\ttzero$ defines 3, in the sense that in every
model \calS\  of $T$ (over a vocabulary with \ttt\ as an identifier),
the interpretation of the identifier \ttt\ is the same
as that of the numeral for 3. Consider instead a 0-ary program defining
an infinite term (i.e.\ essentially a stream),  for instance the
singleton program $I$ consisting of
$\bxtt{ind} = \tts(\bxtt{ind})$.
This does not have any solution in the free algebra
of the unary numerals, that is: the free algebra cannot be expanded into
the richer vocabulary with $\bxtt{ind}$ as a new identifier, so as to satisfy
the equation $I$.\footnote{As usual, when 
a structure is an expansion of another
they have the same universe.}  But $I$ is modeled in any structure
where \tts\ is interpreted as identity, and \bxtt{ind} as any 
structure element.  Thus the interpretation of $\bxtt{ind}$ is not unique.
For a more interesting example, consider the structure consisting of
countable ordinals, with \tts\ interpreted as the function $\grl x. 1+x$.
Then $I$ holds whenever \bxtt{ind} is interpreted as an infinite ordinal.

It follows that in our context bi-simulation, while guaranteeing true equality 
for the canonical model, implies in general
only equivalent computational behavior.
Indeed, in the global semantic context bi-simulation is not a sound inference rule, 
since for example two distinct objects can unfold to exactly the same 
stream of digits (i.e.\ be observationally equivalent).
However, bi-simulation leads to an equivalence relation, which can be captured
by a function $\bxit{bsm}$.  Consider the program consisting of the two
equations
$\bfb(0:x,0:y) = 0: \bfb(x,y)$
and
$\bfb(1:x,1:y) = 1: \bfb(x,y)$.  If $P$ also defines constant identifiers \tta\ and \ttb\ 
as some streams, then we have
$P \models S(\tta) \wedge S(\ttb) \ra S(\bfb(\tta,\ttb))$ 
just in case there is a bi-simulation between the streams denoted 
by \tta\ and \ttb,
i.e.\ they are equal as elements of the coalgebra of boolean streams.
If the equality $\tta=\ttb$ is provable using the traditional coinduction rule for bi-simulation
then the implication $(P) \ra S(\bfb(\tta,\ttb))$ is provable in our deductive calculus
below.
Thus our framework supports all common forms of reasoning about 
coinductive data.

\subsection{Semantics of programs}

The global semantic approach to equational programs, considered for 
inductive data in \cite{Leivant-intrinsic}, is of interest as an alternative 
alternative to the ``canonical-structure" approach.
Under the global semantics approach the notion of 
{\em correctness} of programs
is simple, direct, and informative.
Here a program over inductive data is said to be 
{\em correct} if it 
maps, in every structure, inductive data to inductive data.
This turns out to be equivalent to the program termination (for all input)
in the intended structure (e.g.\ \dN\ when the constructors are \ttzero\ and \tts).
For programs over co-inductive data, which we address here, correctness
will turn out to be equivalent to {\em productivity} (sometimes dubbed 
{\em fairness}): if the input is a stream, then
the program will have a stream as output, without stalling.


The semantics of equational programs
for inductive data, such as the natural numbers, is straightforward.
Given a structure  \calS\ (for a vocabulary including the constructors
in hand),
a program $(P,\bff)$ (unary say) computes the partial 
function $g: \; \dN \pa \dN$  given by: $g(n) = m$ iff
$P \vdash \bff(\bar{n}) = \bar{m}$,
i.e.\ the equation is deducible from $P$ in equational logic. 
(We write $\bar{n}$ for the $n$'th unary numeral $\tts^{[n]}(\ttzero)$.)

Let \calS\ be a structure whose vocabulary contains at least the
constructors in hand.
Consider fresh 0-ary identifiers $\ttv_a$, one for each 
$a \in |\calS|$ (i.e.\ element of the universe of \calS).
In keeping with the terminology of Model Theory,
we define the {\em diagram} of \calS\ to be the theory\fn{We write 
${\bf c}_{\cal S}$ for the interpretation of the identifier 
{\bf c} in the structure $\cal S$.}
$$
\begin{array}{ll}
\bxrm{Diag}(\calS) \quad 
	= & \{ \ttv_a = \bfc( \ttv_{b_1} \cdots \ttv_{b_r}) \; \mid
	\\
	& \qquad\qquad
			a = \bfc_{{\cal S}} (b_1 \cdots b_r)
			\quad \hbox{\bfc\ an $r$-ary constructor } \}
\end{array}
$$

In the presence of coinductive data-types, data may be infinite,
and so the 
operational semantics of equational programs must compute the output
piecemeal from finite information about the input.
If \grG\ is any set of equations,
and $\bft$ and $\bft'$ are terms, we write $\grG \vdash^\omega \bft=\bft'$
if for all deep-destructors $\grP$ we have
(in equational logic)  $\grG, \; \bxrm{Diag} \vdash \grd((\grP(\bft),\vec{x}) = \grd((\grP(\bft'))$.
That is, one can establish equationally the observational equivalence
of \bft\ and $\bft'$, i.e.\ the stepwise equality of
finite approximations of the two terms.

If $\bft'$ is a data term, then $\grG \vdash^\omega \bft=\bft'$ is
clearly equivalent
(by discourse-level induction on $|\bft'|$) to 
$\grG , \; \bxrm{Diag} \vdash \bft = \bft'$.

We say that a $k$-ary program $(P,\bff)$
{\em computes over} \calS\ the partial-function\\
$f: \, |\calS|^k \pa |\calS|$ when
%
for every $\vec{a}, b \in |\calS|$ we have $f(\vec{a}) = b$ just in case
$P \cup \bxrm{Diag}(\calS) \vdash^\omega \bff(\ttv_{a})= \ttv_b$.


\medskip

\noindent
{\bf Examples.}
Consider as constructors two unary functions
(``successors") \ttzero\ and \ttone.
Let \calS\ be the structure of the $\grw$-words over $\{0,1\}$,
with the obvious interpretation of the constructors.
Writing $a$ for $(01)^\omega$ and $b$ for $(10)^\omega$, the diagram of
\calS\ includes the equations $v_a = 0v_b$, and $v_b=1v_a$.
In this simple case these equations could be used to define $a$ and $b$,
but if $c$ and $d$ are the binary expansions of 
$\grp/4$ and $(\grp-2)/2$, then the equation $v_c = 1v_d$ is also
in the diagram, with not much to say about what $c$ and $d$ really are.

The unary program consisting of the two equations
$\bff(0w) = 1\, \bff(w)$,
$\bff(1w) = 0\, \bff(w)$ defines the
function $\bxit{flip}: \; |\calS| \ra  |\calS|$.
We have $\bxit{flip}((01)^\omega ) = (10)^\omega$, because
we can easily see that
$$
P ,\; v_a = 0v_b, \;  v_b=1v_a \vdash^\omega 
	\bxit{flip}(v_a) = v_b
$$
We also have for $e = $ the digitwise flip of $c$ above that
$$ 
P ,\; \bxrm{Diag}(\calS) \vdash^\omega
        \bxit{flip}(c) = e
$$
However, as we take deeper destructors for the two terms,
the equational proof needed here will use increasingly
large (albeit finite) portions of $\bxrm{Diag}(\calS)$.


\subsection{Data systems}

So far we have considered abstract structures, with no {\em a priori}
restriction on the behavior of constructor-identifiers.
We now proceed to define data-types, needed to 
reflect the intended computational behavior
of programs. We use reserved
relation-identifiers (i.e.\ predicate symbols) for data-types,
and convey their defining properties 
by axioms (closure conditions) rather than via \grm\ and \grn\ fixpoint 
operators.
This allows us to incorporate data types seamlessly into the 
(first order) deductive machinery.

Descriptive and deductive tools for inductive and coinductive data
are not new, of course.
For instance, the Common Algebraic Specification Language 
\bxsc{Casl} has been used as a unifying standard in the algebraic 
specification community, and extended to coalgebraic data
\cite{Reichel99,RotheTJ01,MossakowskiSRR06,Schroder08}.
Several frameworks combining inductive and coinductive data,
such as \cite{Padawitz00}, strive to be comprehensive, including
various syntactic distinctions and categories, whereas our approach
is minimalist.  Such minimalism is made possible by combining the
global semantic approach with a semantic (i.e.\ Curry-style)
view of types, by which types indicate semantic properties of 
pre-existing objects, as opposed to the ontological (Church-style) view, 
by which types precede objects, with each object 
coming with a pre-assigned type.

Let $\calC= \{\bfc_1 , \ldots , \bfc_k\}$  be a set of constructors as above,
where $\bfc_i$ is of arity $r_i = \arity(\bfc_i)$.
A {\em data-system} over \calC\ consists of
\be
\item
A list $D_1 \ldots D_k$ (the order matters)
of unary relation-identifiers, where
each $D_n$ is designated as either
an {\em inductive-predicate} or a {\em coinductive-predicate,}
and associated a set $\calC_n \subseteq \calC$ of constructors.
\item
For each constructor \bfc, of arity $r$ say, a non-empty finite set of 
{\em functional types \grt,}
each of the form $E_1 \times \cdot\cdot\cdot \times E_r \ra E_0$,
where each $E_i$ is one of the $D_j$'s.
Here we require that no $E_i$ comes after $E_0$ in the given listing of the
predicates $D_i$.
We say then that \bfc\ {\em has type \grt}.
\ee
The data-systems defined above do not accommodate simultaneous inductive or coinductive
definitions, but a straightforward generalization does. 

\medskip

\noindent
{\bf Example.}
Let \calC\ consist of the identifiers
\bxtt{0, 1, [], s, t,} and \ttc, of arities 0,0,0,1,1, and 2,
respectively.
Consider the following (ordered)
list of predicates: inductive predicate $B$ (for booleans) and $N$ (natural numbers),
coinductive predicates
$J$ (infinite \bxtt{s/t}-words) and $S$ (streams of natural numbers),
and an inductive predicate $L$ (lists of such streams).

The association of types to constructors is as follows. 
$$
\begin{array}{l}
\ttzero:B \quad \ttzero:N\\
\ttone:B\\
\zero []:L\\
\tts: N\sra N \quad \tts:J\sra J\\
\ttt: J\sra J\\
\ttc: N\times S \sra S\\
\ttc: S\times L \sra L
\end{array}
$$

Note that constructors are being reused for different
data-types.
This is in agreement with our untyped, generic approach, where the
intended type information is conveyed by the data-predicates.
In other words, data-types are explicitly conveyed in the
formalism's syntax as semantic (Curry style) rather than
onthological (Church style) properties.
\qed

\medskip


The {\em canonical model} $\calA = \lsem \calD\rsem$ of a data-system
\calD\ consists of interpretations $\lsem D_n \rsem$ ($n=1..k$) of the
data-predicates as sets of finite and infinite terms, obtained
by discourse-level recurrence, as follows.
If $D_n$ is inductive, then $\lsem D_n \rsem$ is the set of terms
obtained from $\lsem D_1\rsem\; \ldots\; \lsem D_{n-1}\rsem$ by 
a finite number of application of the constructors in $\calC_n$;
dually, if $D_n$ is coinductive, then
$\lsem D_n \rsem$ is the set of finite and infinite terms
obtained from $\lsem D_1\rsem\; \ldots\; \lsem D_{n-1}\rsem$ by such
applications.  These terms are trees labeled by constructors, where any node labeled
by a constructor of arity $r$ has $r$ children.
Note that if the (non-empty) set $\calC_n$ of constructors associated with $D_n$ 
has no 0-ary constructors,
then for an inductive $D_n$ the set $\lsem D_n \rsem$ is empty,
whereas for a coinductive $D_n$ it is a nonempty set of
infinite terms.



\nsubsection{Adequacy of Global semantics}

Herbrand famously proposed to define the computable functions 
(over \dN) as those
that are unique solutions of equational programs.  
That definition yields in fact
all the hyper-arithmetical functions, a far larger class.  But Herbrand was not
far off: he only needed to adopt a global approach, rather than restrict
attention to the standard structure of the natural numbers. Indeed,
in \cite{Leivant-intrinsic} we observed the following. 
We say that a structure is {\em data-correct for \dN}
if it interprets the identifier \ttN\ as the set of numeral denotations.




\begin{thm}\label{thm:adequacy}{(\rm Semantic Adequacy Theorem for Inductive Data)}
An equational program $(P,\bff)$ over \dN\ computes a total function
iff the formula $\ttN(x) \ra \ttN(\bff(x))$ is true in every model
of $P$ which is data-correct for \ttN.
\end{thm}

The proof in \cite{Leivant-intrinsic} of the
nontrivial direction of Theorem \ref{thm:adequacy} proceeds by 
constructing a ``test-model" for the program $P$.
One starts with an extended term model, using the program-functions
in $P$ as well the constructors, and takes the
quotient of that term model over the equivalence relation of 
equality-derived-from $P$.

\section{Intrinsic Theories}

Intrinsic theories, introduced in \cite{Leivant-LCC,Leivant-intrinsic} 
for inductive data, are skeletal first-order theories whose interest
lies in a natural and streamlined formalization of reasoning about equational
computing.  For example, the intrinsic theory for the natural numbers 
is suited for incorporating equational programs as axioms,
and while it has the same 
provably computable functions as Peano's Arithmetic, it has a
more immediate formalization of the notion of provable computability.
For background, rationale, and examples, we refer to \cite{Leivant-intrinsic}.

\medskip

The {\em intrinsic theory} for a data-system \calD, $\bxbf{IT}(\calD)$, has
\bi
\item
The rules of \calD; 
\item
{\em Injectiveness axioms} stating that the constructors are injective, i.e.\ for each
$\bfc \in \calC$, of arity $r$,
$$
\forall x_1 \ldots x_r, \,y_1 \ldots y_r \;\;
	\bfc(\vec{x})=\bfc(\vec{y}) \ra \bigwedge_i \; x_i=y_i
$$
\item
{\em Separation axioms} stating that the constructors have disjoint images:
$$
\forall \vec{x},\, \vec{y} \;\;
	\bfc\vec{x} \neq \bfd\vec{y}
$$
for each distinct constructors $\bfc,\bfd$;
and 
\item
For each constructor \bfc, and type $E_1 \times \cdot\cdot\cdot\times E_r \ra E_0$
for \bfc,
with $E_0$ an inductive predicate,
the corresponding clause in the inductive definition of $E_0$.
That is, the {\em data-introduction} rule
$$
\infer{E_0(\bfc\, x_1 \cdots x_r)}
	{E_1(x_1) & \cdot\cdot\cdot & E_r(x_r)}
$$
These rules delineate the intended meaning of $E_0$ from below.
\item
For each constructor \bfc, and type $E_1 \times \cdot\cdot\cdot\times E_r \ra E_0$
for \bfc,
with $E_0$ a {\em co-inductive} predicate,
the corresponding clause in the co-inductive definition of $E_0$.
That is, the {\em data-elimination} rule
$$
\infer{E_i(x_i)}{E_0(\bfc\, x_1 \cdots x_r)}
$$
These rules delineate the intended meaning of a coinductive $E_0$ from above.

\item
For each inductive data-predicate $D_n$ as above, a data-elimination 
(i.e.\ Induction) rule: for each formula\footnote{We use the bracket 
notation $\varphi[t]$ to stand for the correct substitution in $\varphi$ of
$t$ for the free occurrences of some fixed variable $z$.}
$\grf\equiv \grf[z]$, the rule 
$$
\infer{\grf[\bft]}
	{D_n(\bft)
	&
	\cm_n[\grf]}
$$
where
$$
\begin{array}{c}
\cm_n[\grf] \quad \equiv \quad 
\end{array}
\qquad
\begin{array}{c}
\infer{\grf[\bft]}
	{D_n(\bft)
	&
	\hbox{\Large\{} \!\!\!\!\!\!  &
	\deduce{\grf[\bfc( x_1 \cdots x_r)]}
		{\deduce{\cdot}{\deduce{\cdot}{\deduce{\cdot}
			{\{E_1^\varphi(x_1)\} \cdot \cdot \cdot \{E_r^\varphi(x_r)\} }}}}
	& 
	\!\!\!\!\!\!  \hbox{\Large\}}_{c:E_1\times \cdots \times E_r \ra D_n}}
\end{array}
$$
Here
$E_i^\varphi(u)$ is $\grf[u]$ if $E_i$ is $D_n$, and is $E_i(u)$ otherwise.
(These open assumptions are closed by the inference.)

That is, if $\grf[u]$ has the same closure properties under the constructors as $D_n$,
then $D_n(\bft) \sra \grf[\bft]$.
\item
For each coinductive data-predicate $D_n$, a data-introduction
(i.e.\ coinduction) rule: for each formula $\grf[z]$, 
\begin{equation}\label{eq:generic-coind}
\infer{D_n(\bft)}
	{\grf[\bft]
	&
	\dc_n[\grf]}
\end{equation}
where
$$
\begin{array}{c}
\dc_n[\grf] \quad \equiv \quad
\end{array}
\quad
\begin{array}{c}
		\deduce{\quad \bigvee\; \{\;
			\exists z_1 \ldots z_r . (\wedge_i E_i^\varphi(z_i) )
				\;\wedge\; x = \bfc(\vec{z}) \mid
				\bfc:E_1 \times \cdots\ \times E_r \ra D_n   \}}
		{\deduce{\cdot}{\deduce{\cdot}{\deduce{\cdot}{\{\grf[x]\}}}}}
\end{array}
$$
(Here $Q_i^{\varphi}$ is defined as for the induction template above.)

That is, if \grf\ has the same closure properties under data decomposition
(i.e.\ the destructors) as $D_n$, then $\grf[\bft] \ra D_n(\bft)$.
\ei

\noindent
{\bf Note.}
Since our approach here is generic to all structures, 
the bounding condition in the statement of Coinduction
is necessary.  Consider for example the coinductive data $W^\infty$
of infinite 0-1 words, i.e.\ the coinductive data predicate built from 
unary function identifiers \ttzero\ and \ttone, considered above.
Taking the eigen formula \grf\ of Coinduction to be $x=x$,
we would get, absent the bounding condition, $\forall x \; W^\infty(x)$,
which is not valid in models of the intrinsic theory for $W$.

From the injectiveness and separation axioms it follows that
it is innocuous to use identifiers for destructors and discriminator functions, as above.

Theorem \ref{thm:adequacy} justifies
a concept of {\em provable} correctness of programs:
$(P,\bff)$ is provably correct in a given formal theory
if the formula above is not merely true in all data-correct models of $P$, but
is indeed provable in the intrinsic theory $\bxbf{IT}(\calD)$
from (the universal closure of) $P$, as an axiom.


\section{Corecurrence and strictly-positive coinduction}

\subsection{Functions definition by corecurrence}

A function definition by recurrence uses its input by
eager evaluation:
it consumes the top constructor of the input to select
the definition-case,
and proceeds to consume that constructor's arguments.
That is, for each constructor \bfc, one has a clause
\begin{equation}\label{eq:recurrence}
\begin{array}{rclc}
f(\bfc(x_1 \ldots x_r),\vec{y})
	&=& g_c(e_1 \ldots e_r,\vec{y}) \quad 
	& r = \underline{\rm arity}(\bfc) \quad e_i \df f(x_i,\vec{y})
\end{array}
\end{equation}
Here each $g_c$ is a previously defined function of appropriate arity.
Using a discriminator \bxit{case} function, the template above
can be summarized as
\beqnas
f(x,\vec{y}) & = & \bxit{case}(x,e_1 \ldots e_k) \\
		&& \quad e_i \df f(\grp_i(x),\vec{x})
\eeqnas
(Recall that $\grp_i$ is the $i$'th destructor.)


Dually, a definition by {\em corecurrence}
builds up the output: it produces the top constructor of the output,
and proceeds to produce that constructor's arguments:

\begin{equation}\label{eq:corecurrence}
\begin{array}{l}
f(\vec{x})
        = c_{h}(\vec{x},e_1 \ldots e_r) \qquad \quad
	\begin{array}{rcl}
        r &=& \underline{\rm arity}(h(\vec{x}))\\
        e_i &\df& f(\vec{g}_i(\vec{x}))
	\end{array}
\end{array}
\end{equation}
This template can be summarized by 
$$
f(\vec{x}) = \bxit{cocase}(h(\vec{x}),e_1 \ldots e_k) 
	\qquad \quad e_i \df f(\vec{g}_i(\vec{x}))
$$
where $\bxit{cocase}(u,\vec{v})$ returns
the main constructor \bfc\ of $u$,
of arity $r$ say, applied to the first $r$ of the remaining arguments $\vec{v}$.

More generally, we use corecurrence to define as above not a single function $f$, but a
vector $\vec{f}=\lng f_1 \ldots f_k \rng$ of functions:
$$
f_j(\vec{x}) = \bxit{cocase}(h_j(\vec{x}),e_1 \ldots e_k)
        \qquad \quad e_i \df f_{\ell_i}(\vec{g}_{ij}(\vec{x}))
$$

The distinction in (\ref{eq:recurrence}) between the recurrence
argument and the parameters $\vec{y}$ disappears
in (\ref{eq:corecurrence}) because the focus
of the definition shifts to the output, which plays a role analogous to the recurrence
argument of the recurrence schema.

When we have just one constructor, e.g.\ 
a binary function \bxit{cons}, the output's main constructor need not be
specified, and (\ref{eq:corecurrence}) can be conveyed by applying 
destructors to the output:
\begin{eqnarray}\label{eq:stream-corecurrence}
\grp_i (f(\vec{x})) &=& f(\vec{g}_i(\vec{x})) \quad i=0,1
\end{eqnarray}
Such use of destructors is common in presentations of corecurrence,
but it fails to capture corecurrence for arbitrary coinductive data.
Of course, each case can be coded using streams,
just as all inductive data can be coded using the natural numbers.

In our untyped setting the values
$f(\vec{g}_0(\vec{x}))$ and $f(\vec{g}_1(\vec{x}))$ have the same
standing.  Streams 
over a finite base set $A$ can be construed as a restricted form of
(\ref{eq:corecurrence}), with each $a \in A$ taken as 
a nullary constructor, and requiring the first argument of \bxit{cons} to be
one of these constructors.

A function over the given data-system is {\em primitive corecursive}
if it is generated from the constructors and destructors
by composition and corecurrence.


\renewcommand{\hd}{\bxit{hd}}
\renewcommand{\tl}{\bxit{tl}}

\bigskip

\noindent
{\bf Example.}  Boolean streams form a simple data system of the kind mentioned
above: {\sc cons} is the unique non-constant constructor, which we denote by an infixed
colon.  The remaining constructs are the nullary \bfzero\ and \bfone,
and the data-predicates are the inductive (and finite) $B$ (booleans) 
and the coinductive $S$ (streams). 
The rules are
$$
\begin{array}{c}
\infer{B(\bfzero)}{}
\end{array}
\quad
\begin{array}{c}
\infer{B(\bfone)}{}
\end{array}
\qquad
\begin{array}{c}
\infer{B(x)}{S(x:y)}
\end{array}
\quad
\begin{array}{c}
\infer{S(y)}{S(x:y)}
\end{array}
$$
The constructor {\it cons} has the
the two destructors $\hd: \,S \ra B$ and $\tl:\, S \ra S$.

Since there is a single non-constant constructor here, corecursion
can be formulated using the destructors, as in the template:
\beqnas
\hd(f(x,\vec{y})) &=& g_0(x,\vec{y})\\ 
\tl(f(x,\vec{y})) &=& f(g_1(x,\vec{y}),\vec{y})
\eeqnas

For example, we can define by corecurrence a function $\bxit{even}$:
$$
\hd(\bxit{even}(x)) = \hd(x);  \quad
	\tl(\bxit{even}(x)) = \bxit{even}(\tl(\tl(x))).  
$$
The function \bxit{even} is productive (i.e.\ fair, see \cite{Sijtsma89,EndrullisGHIK07}), 
in the sense that it maps streams to streams. 

More precisely,
in every model \calS\ of the data-system, expanded to interpret
\bxit{even} while satisfying its equational definition,
if $S(x)$ holds for $x$ bound to an element $a$ of \calS's universe, then $S(\bxit{even}(x))$.

The generic coinduction rule (\ref{eq:generic-coind}) specializes for boolean streams
to the following.
\begin{equation}\label{eq:stream-coind}
\begin{array}{c}
\infer{S(\bft)}
	{\grf[\bft]
	&
		\deduce{\exists z_0,z_1 . (B(z_0) \wedge \grf[z_1]
				\wedge x = z_0:z_1}
		{\deduce{\cdot}{\deduce{\cdot}{\deduce{\cdot}{\deduce{\zero}
			{\{\grf[x]\}}}}}}}
\end{array}
\end{equation}

While corecurrence is dual to recurrence, it is computationally weaker in some ways.
Recurrence allows coding of computation traces, so that cumulative (course-of-value) recurrence
is implementable using simple recurrence.
In contrast, a cumulative variant of corecursion, using at any given point the output stream
so far, is not captured by standard corecurrence.
For example, the definition of the Morse-Thue sequence,
$x=1:\bxit{merge}(x, not(x))$, is not a legal corecurrence.

\subsection{Strictly-positive coinduction captures corecurrence}

Consider the intrinsic theory for a coinductive datatype, such as the boolean streams.
We call a formula {\em strongly positive}
if built using conjunction, disjunction, and $\exists$ as the only logical operations.
A formula is {\em unipolar} if it does not have both positive
and negative occurrences of data-predicates.
As mentioned in the Introduction above,
we know that a function over \dN\ is primitive recursive iff it is provably
correct, using classical logic, in the intrinsic theory for \dN\
with induction restricted to unipolar formulas;
and also iff it is provably
correct, using minimal logic, in the intrinsic theory for \dN\
with induction restricted to strongly-positive formulas.

\newcommand{\strm}{{\calS}{\large\it m}}
\renewcommand{\itp}{\hbox{$\bxbf{IT}^+$}}
\newcommand{\itpp}{\hbox{$\bxbf{IT}^+(P)$}}

Here we prove for the primitive corecursive functions an analog of the latter characterization.
For concreteness and expository economy, we focus on the data-system \strm\
consisting of just streams of booleans as data-type, and refer to 
the intrinsic theory for it,
based on minimal logic.  We write $\bxbf{IT}^+$ for that theory, with
coinduction restricted to strictly-positive formulas. 

\begin{prop}\label{prop:corec-to-coind}
If a $k$-ary $f$ is defined by corecursion from functions provable in \itp,
then $f$ is provable in \itp.  
\end{prop}
\prf
Suppose that $f$ is defined by
$$
f(x) = g_0(x)\,:\, f(g_1(x))
$$
Let $(P_0,g_0)$ and $(P_1,g_1) $ be programs (with no common
function-identifiers) that are provable in \itp, with
$\calD_0$ a derivation of $B(g_0(u))$ from $S(u)$ and $P_0$, and
$\calD_1$ deriving $S(g_1(u))$ from $S(u)$ and $P_1$.
Consider $(P,f)$ where $P$ is $P_0 \cup P_1$ augmented
with the corecursive definition of $f$ from $g_0$
and $g_1$.
Then $S(f(x))$ is derived from $S(x)$ and $P$, as follows.

Let $\grf[z]$ be the strictly-positive formula
$\exists y\, S(y) \wedge f(y) = z$.
Then $S(f(x))$ is derived from assumptions $S(x)$ and $P$ by
coinduction on \grf, since the premises of coinduction
follow from these assumptions:
\bi
\item
From $S(x)$ we have $S(x) \wedge f(x)=f(x)$, and so $\grf[f(x)]$.
\item
Assuming $\grf[x]$ we have $S(y) \wedge f(y)=x$ for some $y$,
i.e.\ $g_0(y):g_1(y)=x$.  But $S(y)$
implies $B(g_0(y))$  by $\calD_0$, 
and $S(g_1(y))$  by $\calD_1$.  Using $\calD_0$ and $\calD_1$ for $u=g_1(y)$,
we get from $S(g_1(y))$ that $\grf[g_1(y)]$.  

Taking $z_0 = g_0(y)$ and $z_1 = g_1(y)$, we thus have 
$f(x)= z_0 : z_1 \wedge B(z_0) \wedge \grf[z_1]$, concluding the other premise
of the coinduction.
\ei
\qed

\subsection{From coinduction to corecurrence}\label{subsec:derivations}

We proceed to show the converse of Proposition \ref{prop:corec-to-coind},
namely that corecurrence captures strongly-positive coinduction.
If $P$ is an equational program,
let us write \itpp\ for the natural deduction calculus for \itp, augmented with
the program $P$ in the guise of an inference rule:\fn{This deductive style has been used
in research on the Curry-Howard morphism for higher-order logic,
e.g.\ \cite{Leivant-foundational}; it was dubbed ``deduction modulo" in 
\cite{DowekHK03} and subsequent works.}  If $\bft=\bft'$ is an equation in $P$, then
$$
\begin{array}{c}\infer{\gra[\bft]}{\gra[\bft']}\end{array}
\quad \text{and} \quad 
\begin{array}{c} \infer{\gra[\bft']}{\gra[\bft]} \end{array}
$$
are inferences, where \gra\ is any atomic formula.
Clearly, a formula \grf\ is derivable in \itpp\ from assumptions
$\vec{\grq}$ iff \grf\ is derivable in \itp\ from $\vec{\grq}$
plus (the universal closure of) $P$.

A basic observation is the following, where we refer to the usual notion
of logical detours in natural deduction derivations \cite{Prawitz65}.
Recall that a logical detour arises when the major premise of an elimination
rule (for a logical operator) is derived by an introduction rule.

\begin{lem}\label{lem:normalproofs}
1. Every derivation of \itpp\ can be converted to a derivation without logical
detours.\\[1mm]
2.  If \calD\ is a derivation of \itpp\ without logical detours, proving a strongly-positive
formula from strongly-positive assumptions, then every formula in \calD\ is strongly-positive.
\end{lem}
\prf
Part (1) is proved as for first-order logic \cite{Prawitz65}.
Part (2) follows by a straightforward structural induction, 
using the fact that coinduction is restricted to strongly-positive formulas, and that the logic
is minimal.
\qed

\medskip

We define a relation $\calS,\;\grh,\;\grs \Vdash \grf$, i.e.\
{\em the stream \grs\ realizes the formula \grf} in the interpretation
$(\calS,\grh)$ consisting of a model of \itp\ and of $P$, and an environment \grh\
in it.  The definition is by induction on \grf. This relation is defined by
structural recurrence on the formula \grf.  
For a stream $\grs$ we define the streams $\grs_i$ $i\geq 0$ inductively, 
jointly with the streams $\grs_i'$.  The intent is that $\grs_0$
consists of the even-positioned entries of \grs, $\grs_1$ of the even-positioned
entries of the remaining entries, etc.
$\grs_0 = \bxit{even}(\grs)$,
$\grs_0' = \bxit{odd}(\grs)$,
$\grs_{i+1} = \bxit{even}(\grs_i')$,
$\grs_{i+1}' = \bxit{odd}(\grs_i')$.

\bi
\item
$\calS,\;\grh,\;\grs \Vdash S(\bft)$ iff $\grs=\lsem \bft\rsem_{{\cal S}, \eta}X$
and $\grs \in S_{\cal S} $.
\item
$\calS,\;\grh,\;\grs \Vdash \bft = \bft'$ iff 
$\grs=\lsem \bft\rsem_{{\cal S}, \eta}X = \lsem \bft'\rsem_{{\cal S}, \eta}X$.
\item
$\calS,\;\grh,\;\grs \Vdash \grf_0 \wedge \grf_1$
iff $\grs_i \Vdash_{{\cal S}, \eta}X \grf_i$, $i=0,1$.
\item
$\calS\;\grh,\; \grs \Vdash  \grf_0 \vee \grf_1$
iff $\calS<\; \grh,\; \tl \grs \Vdash \grf_{\hd \sigma}$.
\item
$\calS,\;\grh,\; \grs \Vdash \exists x \grf$
iff $\calS,\; \grh [x:= \grs_0], \; \grs_1 \Vdash \grf$.
\ei

\begin{lem}\label{lem:realizability}j
Suppose $\itpp \vdash \wedge_i \, \grq_i[\vec{x}] \sra \grf[\vec{x}]$.
Then there is a primitive corecursive function $f_0$ such that
for all models \calS\ of $P$, and for all streams $\vec{\grt}$ and $\grs_i$, if 
$$
\calS, \; [\vec{x} := \vec{\grt}],\; \grs_i \Vdash \grq_i,
$$
then
$$
\calS, \; [\vec{x} := \vec{\grt}],\; f_0(\vec{\grt},\vec{\grs}) \Vdash \grf.
$$

More precisely, there is a primitive corecursive program $P_0$ 
(which computes $f$ above), such that every model of $P$ can be expanded to a model
of $P_0$, where $f_0$ has the property above.
\end{lem}
\prf
Let $\calD$ be a derivation of $\grq[\vec{x}] \sra \grf[\vec{x}]$ in \itpp.
By Lemma \ref{lem:normalproofs} we may assume that \calD\ is detour-free, and with all
formulas strongly-positive.  The Lemma is proved by structural induction on \calD.
For the base cases $f$ is the identity. The cases where the main inference of \calD\
is a logical rule are immediate from the definition of $\Vdash$.
The cases of Data-elimination rule (decomposition) are immediate since the destructors
functions are initial primitive corecursive functions.
The case of the rewrite rules based on $P$ is assured by the fact that \calS\ is assumed
to be a model of $P$.

The case of interest is where the main inference rule of \calD\ is
Coinduction (for strongly-positive formulas): 
\begin{equation}
\begin{array}{c}
\infer{S(\bft)}
	{\grf[\bft]
	&
		\deduce{\exists z_0,z_1 . (B(z_0) \wedge \grf[z_1]
				\wedge x = z_0\,:\,z_1}
		{\deduce{\cdot\cdot\cdot}
			{\{\grf[x]\}}}}
\end{array}
\end{equation}
By IH applied to the left sub-derivation, there is a primitive corecursive
function $g(\vec{u},\vec{v}))$ 
yielding a stream \grs\ realizing 
$\grf[\bft]$, from an environment $\vec{u}$ and realizers $\vec{v}$ for the open assumptions.
By IH applied to the right sub-derivation, there is a primitive corecursive function
$h(\vec{u},u',\vec{v},v')$ yielding a stream realizing
$$
\grf'[x] \quad :\equiv \quad
	\exists z_0,z_1 . (B(z_0) \wedge \grf[z_1] \wedge x = z_0\,:\,z_1)
$$
from an environment $\vec{u}$, a stream $u'$ assigned to $x$,
realizers $\vec{v}$ for the open assumptions, and a realizer $v'$ for $\grf[x]$ in the environment
$(\vec{v},v')$.  Let $j$ and $j'$ be the functions that extract from a realizer for 
$\grf'$ (in a given environment) the boolean $z_0= \hd(x)$, and the realizer of $z_1=\tl(x)$,
respectively.

If $\vec{u}$ are the variables free in \calD, define by corecurrence
$$
r(\vec{u},\vec{v},w) = j(w)\,:\, r(\vec{u},\vec{v},j'(h(\vec{u},\vec{v},w)))
$$
Thus, if $\vec{u}$ are streams, and $\vec{v}$ are realizers for the open assumptions
of \calD\ for the environment $\vec{u}$, then
$$
r(\vec{u},\vec{v},g(\vec{u},\vec{v}))
$$
is the value of \bft, and therefore a realizer of $S(\bft)$, i.e.\ the conclusion
of \ref{eq:stream-coind}.
\qed

\begin{thm}\label{thm:corec-charact}
A function over streams is primitive corecursive iff it is computable by
some equational program which is provable in \itp.
\end{thm}
\prf
If a function is primitive corecursive then its primitive corecursive definition
is provable in \itp, by Proposition \ref{prop:corec-to-coind}.

Conversely, suppose $f$ is a function computable by some equational programs
$(P,\bff)$ which is provable in \itp, i.e.\ there is a derivation of \itpp\
of the formula $S(x) \ra S(\bff(x))$.
From Lemma \ref{lem:realizability} it follows that there is a primitive corecursive 
program $(P_0,\bff_0)$ such that in all models \calS\ of $P$, a realizer of $S(\grs)$, i.e.\
\grs\ itself, is mapped by $f_0$ to a realizer of $S(\bff(x))$, i.e.\ the value of
$\bff(x)$ in the structure.  Since $f$ is computed by $P$ in the canonical structure,
the above holds there too, i.e.\ $f$ is primitive-corecursive in the canonical structure.
\qed

\bibliographystyle{eptcs}
\bibliography{x}

\begin{thebibliography}{10}
\providecommand{\bibitemdeclare}[2]{}
\providecommand{\urlprefix}{Available at }
\providecommand{\url}[1]{\texttt{#1}}
\providecommand{\href}[2]{\texttt{#2}}
\providecommand{\urlalt}[2]{\href{#1}{#2}}
\providecommand{\doi}[1]{doi:\urlalt{http://dx.doi.org/#1}{#1}}
\providecommand{\bibinfo}[2]{#2}

\bibitemdeclare{article}{AstesianoBKKMST02}
\bibitem{AstesianoBKKMST02}
\bibinfo{author}{Egidio Astesiano}, \bibinfo{author}{Michel Bidoit},
  \bibinfo{author}{H{\'e}l{\`e}ne Kirchner}, \bibinfo{author}{Bernd
  Krieg-Br{\"u}ckner}, \bibinfo{author}{Peter~D. Mosses},
  \bibinfo{author}{Donald Sannella} \& \bibinfo{author}{Andrzej Tarlecki}
  (\bibinfo{year}{2002}): \emph{\bibinfo{title}{{CASL:} the Common Algebraic
  Specification Language}}.
\newblock {\sl \bibinfo{journal}{Theor. Comput. Sci.}}
  \bibinfo{volume}{286}(\bibinfo{number}{2}), pp. \bibinfo{pages}{153--196}.

\bibitemdeclare{article}{BarwiseM78}
\bibitem{BarwiseM78}
\bibinfo{author}{Jon Barwise} \& \bibinfo{author}{Yanis Moschovakis}
  (\bibinfo{year}{1978}): \emph{\bibinfo{title}{Global inductive
  definability}}.
\newblock {\sl \bibinfo{journal}{Journal of Symbolic Logic}}
  \bibinfo{volume}{43}, pp. \bibinfo{pages}{521--534}.

\bibitemdeclare{book}{Buss86}
\bibitem{Buss86}
\bibinfo{author}{Samuel Buss} (\bibinfo{year}{1986}):
  \emph{\bibinfo{title}{Bounded Arithmetic}}.
\newblock \bibinfo{publisher}{Bibliopolis}, \bibinfo{address}{Naples}.

\bibitemdeclare{article}{DowekHK03}
\bibitem{DowekHK03}
\bibinfo{author}{Gilles Dowek}, \bibinfo{author}{Th{\'e}r{\`e}se Hardin} \&
  \bibinfo{author}{Claude Kirchner} (\bibinfo{year}{2003}):
  \emph{\bibinfo{title}{Theorem Proving Modulo}}.
\newblock {\sl \bibinfo{journal}{J. Autom. Reasoning}}
  \bibinfo{volume}{31}(\bibinfo{number}{1}), pp. \bibinfo{pages}{33--72}.

\bibitemdeclare{inproceedings}{EndrullisGHIK07}
\bibitem{EndrullisGHIK07}
\bibinfo{author}{J{\"o}rg Endrullis}, \bibinfo{author}{Clemens Grabmayer},
  \bibinfo{author}{Dimitri Hendriks}, \bibinfo{author}{Ariya Isihara} \&
  \bibinfo{author}{Jan~Willem Klop} (\bibinfo{year}{2007}):
  \emph{\bibinfo{title}{Productivity of Stream Definitions}}.
\newblock In \bibinfo{editor}{Erzs{\'e}bet Csuhaj-Varj{\'u}} \&
  \bibinfo{editor}{Zolt{\'a}n {\'E}sik}, editors: {\sl
  \bibinfo{booktitle}{FCT}}, {\sl \bibinfo{series}{Lecture Notes in Computer
  Science}} \bibinfo{volume}{4639}, \bibinfo{publisher}{Springer}, pp.
  \bibinfo{pages}{274--287}, \doi{10.1007/978-3-540-74240-1\_24}.

\bibitemdeclare{inproceedings}{Fagin74}
\bibitem{Fagin74}
\bibinfo{author}{Ronald Fagin} (\bibinfo{year}{1974}):
  \emph{\bibinfo{title}{Generalized first order spectra and polynomial time
  recognizable sets}}.
\newblock In \bibinfo{editor}{R.~Karp}, editor: {\sl
  \bibinfo{booktitle}{Complexity of Computation}},
  \bibinfo{publisher}{SIAM-AMS}, pp. \bibinfo{pages}{43--73}.

\bibitemdeclare{inproceedings}{Immerman89}
\bibitem{Immerman89}
\bibinfo{author}{Neil Immerman} (\bibinfo{year}{1989}):
  \emph{\bibinfo{title}{Descriptive and Computational Complexity}}.
\newblock In: {\sl \bibinfo{booktitle}{FCT}}, pp. \bibinfo{pages}{244--245}.

\bibitemdeclare{article}{JonesS74}
\bibitem{JonesS74}
\bibinfo{author}{N.G. Jones} \& \bibinfo{author}{A.L. Selman}
  (\bibinfo{year}{1974}): \emph{\bibinfo{title}{Turing machines and the spectra
  of first-order formulas}}.
\newblock {\sl \bibinfo{journal}{Journal of Symbolic Logic}}
  \bibinfo{volume}{39}, pp. \bibinfo{pages}{139--150}.

\bibitemdeclare{book}{Kleene69}
\bibitem{Kleene69}
\bibinfo{author}{Stephen~C.\ Kleene} (\bibinfo{year}{1969}):
  \emph{\bibinfo{title}{Formalized Recursive Functions and Formalized
  Realizability}}.
\newblock {\sl \bibinfo{series}{Memoirs of the AMS}}~\bibinfo{volume}{89},
  \bibinfo{publisher}{American Mathematical Society},
  \bibinfo{address}{Providence}.

\bibitemdeclare{article}{Leivant-foundational}
\bibitem{Leivant-foundational}
\bibinfo{author}{Daniel Leivant} (\bibinfo{year}{1994}):
  \emph{\bibinfo{title}{A foundational delineation of poly-time}}.
\newblock {\sl \bibinfo{journal}{Information and Computation}}
  \bibinfo{volume}{110}, pp. \bibinfo{pages}{391--420}.

\bibitemdeclare{inproceedings}{Leivant-LCC}
\bibitem{Leivant-LCC}
\bibinfo{author}{Daniel Leivant} (\bibinfo{year}{1995}):
  \emph{\bibinfo{title}{Intrinsic theories and computational complexity}}.
\newblock In \bibinfo{editor}{D.~Leivant}, editor: {\sl
  \bibinfo{booktitle}{Logic and Computational Complexity}},
  \bibinfo{series}{LNCS}, \bibinfo{publisher}{Springer-Verlag},
  \bibinfo{address}{Berlin}, pp. \bibinfo{pages}{177--194}.

\bibitemdeclare{article}{Leivant-intrinsic}
\bibitem{Leivant-intrinsic}
\bibinfo{author}{Daniel Leivant} (\bibinfo{year}{2002}):
  \emph{\bibinfo{title}{Intrinsic reasoning about functional programs {I}:
  {F}irst order theories}}.
\newblock {\sl \bibinfo{journal}{Annals of Pure and Applied Logic}}
  \bibinfo{volume}{114}, pp. \bibinfo{pages}{117--153},
  \doi{10.1016/S0168-0072(01)00078-1}.

\bibitemdeclare{article}{Leivant-unipolar}
\bibitem{Leivant-unipolar}
\bibinfo{author}{Daniel Leivant} (\bibinfo{year}{2004}):
  \emph{\bibinfo{title}{Intrinsic reasoning about functional programs {II}:
  unipolar induction and primitive-recursion}}.
\newblock {\sl \bibinfo{journal}{Theor. Comput. Sci.}}
  \bibinfo{volume}{318}(\bibinfo{number}{1-2}), pp. \bibinfo{pages}{181--196},
  \doi{10.1016/j.tcs.2003.11.002}.

\bibitemdeclare{article}{Moschovakis89}
\bibitem{Moschovakis89}
\bibinfo{author}{Yiannis~N. Moschovakis} (\bibinfo{year}{1989}):
  \emph{\bibinfo{title}{The Formal Language of Recursion}}.
\newblock {\sl \bibinfo{journal}{J. Symb. Log.}}
  \bibinfo{volume}{54}(\bibinfo{number}{4}), pp. \bibinfo{pages}{1216--1252},
  \doi{10.2307/2274814}.

\bibitemdeclare{article}{MossakowskiSRR06}
\bibitem{MossakowskiSRR06}
\bibinfo{author}{Till Mossakowski}, \bibinfo{author}{Lutz Schr{\"o}der},
  \bibinfo{author}{Markus Roggenbach} \& \bibinfo{author}{Horst Reichel}
  (\bibinfo{year}{2006}): \emph{\bibinfo{title}{Algebraic-coalgebraic
  specification in {CoCasl}}}.
\newblock {\sl \bibinfo{journal}{J. Log. Algebr. Program.}}
  \bibinfo{volume}{67}(\bibinfo{number}{1-2}), pp. \bibinfo{pages}{146--197},
  \doi{10.1016/j.jlap.2005.09.006}.
\newblock \urlprefix\url{http://dx.doi.org/10.1016/j.jlap.2005.09.006}.

\bibitemdeclare{book}{Mosses04}
\bibitem{Mosses04}
\bibinfo{author}{Peter~D. Mosses} (\bibinfo{year}{2004}):
  \emph{\bibinfo{title}{CASL Reference Manual, The Complete Documentation of
  the Common Algebraic Specification Language}}.
\newblock {\sl \bibinfo{series}{Lecture Notes in Computer Science}}
  \bibinfo{volume}{2960}, \bibinfo{publisher}{Springer}, \doi{10.1007/b96103}.

\bibitemdeclare{article}{Padawitz00}
\bibitem{Padawitz00}
\bibinfo{author}{Peter Padawitz} (\bibinfo{year}{2000}):
  \emph{\bibinfo{title}{Swinging types=functions+relations+transition
  systems}}.
\newblock {\sl \bibinfo{journal}{Theor. Comput. Sci.}}
  \bibinfo{volume}{243}(\bibinfo{number}{1-2}), pp. \bibinfo{pages}{93--165},
  \doi{10.1016/S0304-3975(00)00171-7}.

\bibitemdeclare{incollection}{Parsons70}
\bibitem{Parsons70}
\bibinfo{author}{Charles Parsons} (\bibinfo{year}{1970}):
  \emph{\bibinfo{title}{On a number-theoretic choice schema and its relation to
  induction}}.
\newblock In \bibinfo{editor}{A.\ Kino}, \bibinfo{editor}{J.\ Myhill} \&
  \bibinfo{editor}{R.\ Vesley}, editors: {\sl \bibinfo{booktitle}{Intuitionism
  and Proof Theory}}, \bibinfo{publisher}{North-Holland},
  \bibinfo{address}{Amsterdam}, pp. \bibinfo{pages}{459--473},
  \doi{10.1016/S0049-237X(08)70771-7}.

\bibitemdeclare{book}{Prawitz65}
\bibitem{Prawitz65}
\bibinfo{author}{D.~Prawitz} (\bibinfo{year}{1965}):
  \emph{\bibinfo{title}{Natural Deduction}}.
\newblock \bibinfo{publisher}{Almqvist and Wiksell},
  \bibinfo{address}{Uppsala}.

\bibitemdeclare{inproceedings}{Reichel99}
\bibitem{Reichel99}
\bibinfo{author}{Horst Reichel} (\bibinfo{year}{1999}): \emph{\bibinfo{title}{A
  Uniform Model Theory for the Specification of Data and Process Types}}.
\newblock In \bibinfo{editor}{Didier Bert}, \bibinfo{editor}{Christine Choppy}
  \& \bibinfo{editor}{Peter~D. Mosses}, editors: {\sl
  \bibinfo{booktitle}{WADT}}, {\sl \bibinfo{series}{Lecture Notes in Computer
  Science}} \bibinfo{volume}{1827}, \bibinfo{publisher}{Springer}, pp.
  \bibinfo{pages}{348--365}, \doi{10.1007/978-3-540-44616-3\_20}.

\bibitemdeclare{article}{RotheTJ01}
\bibitem{RotheTJ01}
\bibinfo{author}{Jan Rothe}, \bibinfo{author}{Hendrik Tews} \&
  \bibinfo{author}{Bart Jacobs} (\bibinfo{year}{2001}):
  \emph{\bibinfo{title}{The Coalgebraic Class Specification Language {CCSL}}}.
\newblock {\sl \bibinfo{journal}{J. UCS}}
  \bibinfo{volume}{7}(\bibinfo{number}{2}), pp. \bibinfo{pages}{175--193}.
\newblock
  \urlprefix\url{http://www.jucs.org/jucs_7_2/the_coalgebraic_class_specificat%
ion}.

\bibitemdeclare{article}{Schroder08}
\bibitem{Schroder08}
\bibinfo{author}{Lutz Schr{\"o}der} (\bibinfo{year}{2008}):
  \emph{\bibinfo{title}{Bootstrapping Inductive and Coinductive Types in
  {HasCASL}}}.
\newblock {\sl \bibinfo{journal}{Logical Methods in Computer Science}}
  \bibinfo{volume}{4}(\bibinfo{number}{4}), \doi{10.2168/LMCS-4(4:17)2008}.
\newblock \urlprefix\url{http://dx.doi.org/10.2168/LMCS-4(4:17)2008}.

\bibitemdeclare{article}{Sijtsma89}
\bibitem{Sijtsma89}
\bibinfo{author}{Ben~A. Sijtsma} (\bibinfo{year}{1989}):
  \emph{\bibinfo{title}{On the Productivity of Recursive List Definitions}}.
\newblock {\sl \bibinfo{journal}{ACM Trans. Program. Lang. Syst.}}
  \bibinfo{volume}{11}(\bibinfo{number}{4}), pp. \bibinfo{pages}{633--649},
  \doi{10.1145/69558.69563}.

\bibitemdeclare{inproceedings}{Tarski52}
\bibitem{Tarski52}
\bibinfo{author}{Alfred Tarski} (\bibinfo{year}{1952}):
  \emph{\bibinfo{title}{Some notions and methods on the borderline of algebra
  and metamathematics}}.
\newblock In: {\sl \bibinfo{booktitle}{Proceedings of the International
  Congress of Mathematicians I}}, \bibinfo{publisher}{American Mathematical
  Society}, \bibinfo{address}{Providence, RI}, pp. \bibinfo{pages}{705--720}.

\end{thebibliography}

\end{document}